\documentclass[prb,preprint,groupedaddress,english]{revtex4-1}
\usepackage{amsmath}
\usepackage{graphicx}

\begin{document}

\title{Massless Dirac Fermions in Graphene under an External Periodic Magnetic Field}

\author{Shuanglong Liu$^1$}

\author{Argo Nurbawono$^1$}

\author{Na Guo$^1$}

\author{Chun Zhang$^{1,2}$}
\email{phyzc@nus.edu.sg}

 
\affiliation{
    $^1$Department of Physics, National University of Singapore,
        2 Science Drive 3, Singapore 117542\\
    $^2$Department of Chemistry, National University of Singapore, 3 Science Drive 3, Singapore 117543}
    
\date{\today}

\begin{abstract}
By solving two-component spinor equations for massless Dirac Fermions, we show that graphene under a periodic external magnetic field exhibits a unique energy spectrum: At low energies, Dirac Fermions are localized inside the magnetic region with discrete Landau energy levels, while at higher energies, Dirac Fermions are mainly found in non-magnetic regions with continuous energy bands originating from wavefunctions analogous to particle-in-box states of electrons. These findings offer a new methodology for the control and tuning of massless Dirac Fermions in graphene. 
\end{abstract}

\maketitle

Due to the Honeycomb lattice structure, the low-energy quasi-particles in graphene behave as massless Dirac Fermions, which leads to peculiar electronic properties and lots of exciting new phenomena.~\cite{Novoselov, Castro Neto} Recently, controlling and tuning properties of massless Dirac Fermions in graphene by an external field has become a hot subject, which may lead to novel applications of graphene in nanoscale devices. It has been found by theoretical calculations that with an external periodic electric field, interesting phenomena such as anisotropic group velocities
of Dirac fermions~\cite{ParkNP}, emerging zero-energy states~\cite{Fertig09},
new type of massless charge carriers~\cite{ParkPRL},
unusual Landau levels and quantum Hall effects~\cite{ParkQHE}, and bandgap opening or quenching~\cite{Aihua}, were reported. With an external local magnetic field, previous studies suggested that the confinement of massless Dirac Fermions in graphene is possible.\cite{Peres, De Martino, Z.Z.Zhang} In current study, we theoretically investigated the properties of low-energy quasi-particles in graphene under an external periodic magnetic field. Our studies showed that a unique energy spectrum of massless Dirac Fermions occurs due to the external periodic magnetic field, which offers new opportunities for future applications of graphene.

When under an external magnetic field, the properties of massless Dirac Fermions in graphene can be described by a spinor equation~\cite{Castro Neto, Lukose} as shown below (Eq. 1), where $v_{F}$ is the Fermi velocity in graphene, $\boldsymbol{\sigma}=(\sigma_{x},\sigma_{y})$ are two-component Pauli matrices, and $\boldsymbol{A}$ is the vector potential corresponding to the magnetic field that is normal to the graphene ($xy$ plane),
\begin{equation}
	\label{eqn:dirac}
	v_{F}\boldsymbol{\sigma}\cdot(-i{\hbar}\boldsymbol{\nabla}+e\boldsymbol{A})\psi(x,y)=E\psi(x,y).
\end{equation}
In the case of uniform magnetic field, with Landau gauge, the vector potential can be written as $\boldsymbol{A}(x)=xB\boldsymbol{\hat{y}}$, and then Eq. 1 has analytical solutions which reads~\cite{Lukose}
\begin{equation}
	\label{eqn:ue}
	\begin{split}
	&E_{n,k_{y}}=sgn(n){\sqrt{2|n|}}{\frac{{\hbar}v_{F}}{l_{c}}}\\
	&{\psi}_{n,k_{y}}\propto e^{ik_{y}y}\begin{pmatrix}sgn(n){\varphi}_{|n|-1}(\xi)\\i{\varphi}_{|n|}(\xi)\end{pmatrix},
  \end{split}
\end{equation}
where $l_{c}=\sqrt{\hbar/eB}$, $\xi\equiv\frac{1}{l_{c}}(x+{l_{c}^{2}}k_{y})$, $n$ are integers and ${\varphi}_{n}(\xi)$ are harmonic oscillator eigenfunctions. The solutions in Eq. 2 perfectly explains the experimentally observed unusual $\sqrt{|n|}$ dependence of discrete Landau levels in graphene~\cite{Jiangz}. 
 
In this paper, we focused on effects of external periodic magnetic fields on properties of massless Dirac Fermions that have never been discussed before. The external periodic magnetic field can be either one dimensional (1d) or two dimensional (2d). In Fig. 1a, we show an example of 1d magnetic field consisting of alternating magnetic and non-magnetic regions. The corresponding periodic vector potential under the Landau gauge is also plotted in the figure. An example of 2d magnetic fields is shown in Fig. 2b. In both examples, the average magnetic field in one unit cell is zero. The resulted periodic systems is referred to as anti-ferromagnetic (AF) superlattices in this paper. We would like to mention here that the theoretical techniques we will discuss later can also be applied to ferromagnetic superlattices with non-zero average magnetic field, and the major physics presented in the paper remain the same in ferromagnetic cases.

\begin{figure}
	\centering
	\includegraphics[width=12cm]{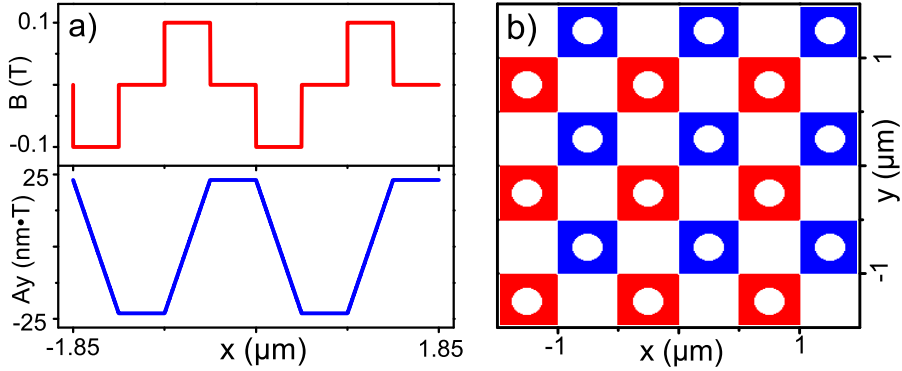}
	\caption{\label{fig:1dmsl} (Color online) Two examples of magnetic superlattices. (a) 1d AF superlattice with alternating magnetic and non-magnetic regions. (b) A 2d AF superlattice. The red (blue) color denotes magnetic field of 0.2 (-0.2) T, and the white color represents non-magnetic regions.}
\end{figure}

Here we describe the theoretical method of solving the spinor equation (Eq. 1) with an external periodic magnetic field. In this case, the solution takes the format of Bloch function, $\psi(x,y)=e^{i(k_{x}x+k_{y}y)}\left(\begin{smallmatrix}\phi_{1}(x,y)\\\phi_{2}(x,y)\end{smallmatrix}\right)$, where $\phi_{1,2}(x,y)$ are periodic. The discrete Fourier expansion can then be applied:
\begin{eqnarray}
	\phi_{1}(x,y) &=& \overset{N}{\underset{m,n=-N}{\sum}}a_{mn}e^{im\omega_{x}x}e^{in\omega_{y}y},\nonumber \\
	\phi_{2}(x,y) &=& \overset{N}{\underset{m,n=-N}{\sum}}b_{mn}e^{im\omega_{x}x}e^{in\omega_{y}y}, \nonumber
\end{eqnarray}
where $\omega_{x}=\frac{2\pi}{T_{x}}$, $\omega_{y}=\frac{2\pi}{T_{y}}$ with $T_{x}$ and $T_{y}$ the periodicity of the superlattice. In our calculations, for 2d case, we use the symmetric gauge, and for 1d superlattice, we use the landau gauge with $T_{y}$ set to be infinite.

Insert the above Fourier expansion into Eq. 1, and using the orthogonality condition, we obtain an eigenvalue equation in matrix form that can be numerically solved,
\begin{equation}
	\left(\begin{array}{cc}0 & h\\h^{\dagger} & 0\end{array}\right)
	\left(\begin{array}{c}a\\b\end{array}\right)
	=E\left(\begin{array}{c}a\\b\end{array}\right)
\end{equation}
where 
\begin{eqnarray}
	a &=& \begin{pmatrix}a_{00} ... a_{2N+1,2N+1}\end{pmatrix}^{T}, \nonumber \\
	b &=& \begin{pmatrix}b_{00} ... b_{2N+1,2N+1}\end{pmatrix}^{T}, \nonumber
\end{eqnarray}
and
\begin{eqnarray}
	h_{m^{\prime}n^{\prime}mn}
	&=&(m^{\prime}\omega_{x}-in^{\prime}\omega_{y}+k_{x}-ik_{y})\delta_{m^{\prime}m}\delta_{n^{\prime}n} \nonumber \\
	&&+A_{xm^{\prime}n^{\prime}mn}-iA_{ym^{\prime}n^{\prime}mn}, \nonumber \\
	A_{x(y)m^{\prime}n^{\prime}mn}
	&=&\frac{1}{T_{x}T_{y}}\iint dydxe^{-i(m^{\prime}\omega_{x}x+n^{\prime}\omega_{y}y)}\cdot \nonumber \\
	&&A_{x(y)}(x,y)\cdot e^{i(m\omega_{x}x+n\omega_{y}y)}. \nonumber
\end{eqnarray}
Eq. 3 is the key equation of this paper, providing the theoretical basis for understanding properties of massless Dirac Fermions under external periodic magnetic field with different gauges. We now solve this equation for different magnetic superlattices. 

\begin{figure}
	\centering
	\includegraphics[clip,width=16cm]{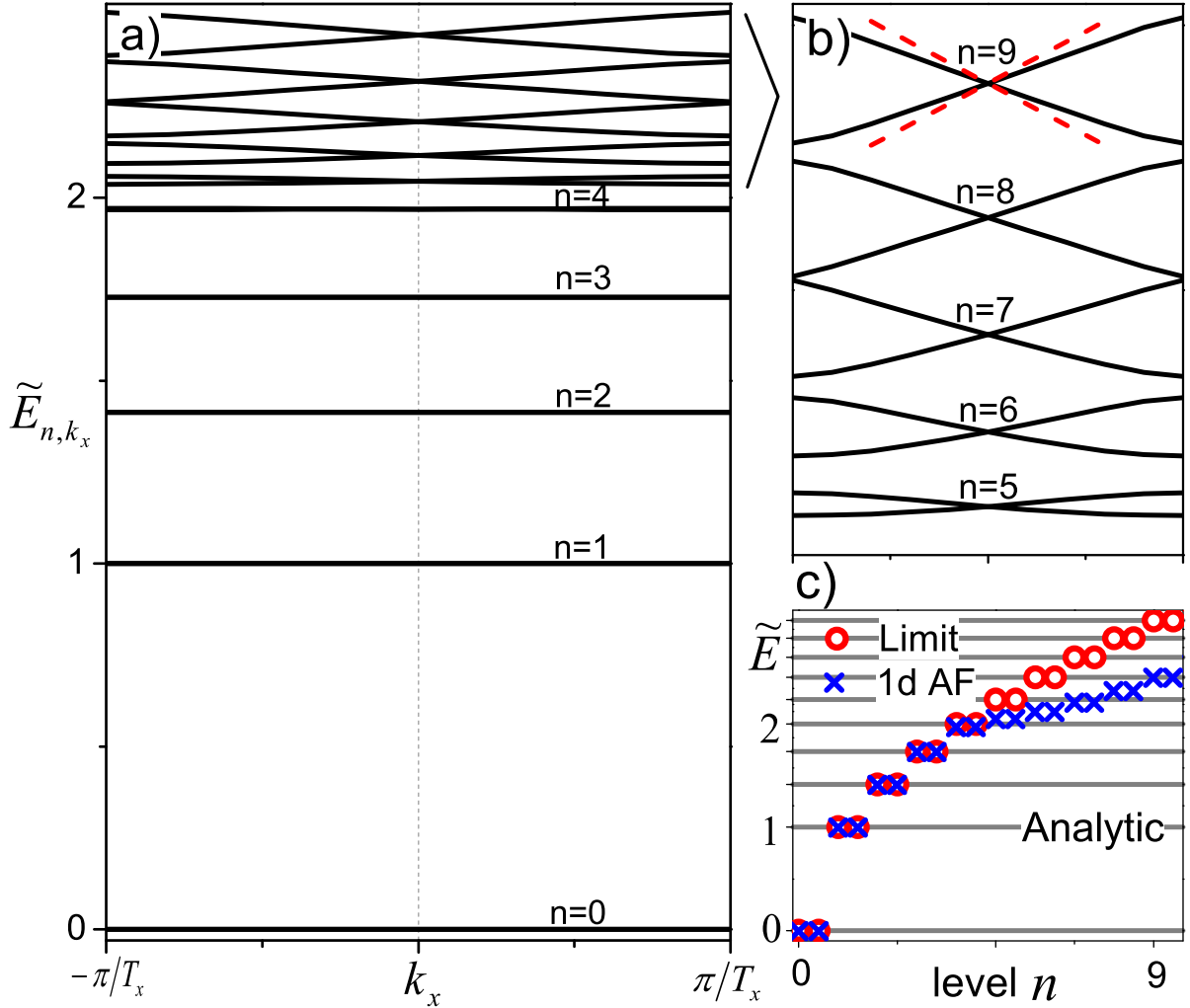} \caption{\label{fig:afew} (Color online) Energy spectrum and wavefunctions of massless Dirac Fermions in 1d magnetic superlattice as depicted in Fig. 1a. (a) The energy spectrum: Discrete Landau levels at lower energies and continuous energy bands at higher energies. (b) Enlarged picture for continuous spectrum. The energy dispersion of free Dirac Fermions (dotted line) is also plotted for comparison. (c) The wavefunction derived from n=5 Landau level. (d) The wavefunction derived from n=6 Landau level. Inset I: Energies calculated at $\Gamma$ point (the center of Brillouin zone). The calculated Landau energy levels under a limiting case with large lattice constant and magnetic regions (red circle) is also plotted for comparison. In this case, the calculated energies is almost exactly the same as analytic solutions shown in Eq. 1 (horizontal dotted lines). Inset II: The wavefunction of n=4 Landau level.}
\end{figure}

In Fig. 2, we show the calculated energy spectrum and wavefunctions at different energies of Dirac Fermions for the 1d AF superlattice depicted in Fig. 1a. As a test of our calculations, we first calculated the energy spectrum for a limiting case with very large size of magnetic regions (1250 $nm$) and also lattice constant (5000 $nm$). This limiting case can be compared with the graphene under a uniform magnetic field. The calculated energy as a function of level index is shown in Inset I (red circle) of the figure. For comparison, analytic Landau energy levels (Eq. 2) with the same magnetic field ($B= 0.1 T$) are also plotted in the inset (dotted lines). Clearly, our numerical calculations exactly reproduced the well known Landau levels of graphene under uniform field. Note that in our case, each of these Landau levels is double degenerate. The double degeneracy comes from the fact that in the AF superlattice we studied, there are two magnetic traps with opposite magnetic fields in one unit cell. We then set the lattice constant to 1850 $nm$, and the size of magnetic region to 462.5 $nm$. Under the same magnetic field, the calculated energy dispersion in the Brillouin zone is plotted in Fig. 2a, and the energies at $\Gamma$ point is shown in Inset I. At low energies (less than the energy of n=4 Landau level in the uniform field), the energy spectrum of the AF superlattice is the same as the well-known discrete Landau levels in graphene. Starting from n=5, the Landau level develops into continuous band which can be seen from Fig. 1a. The transition of the energy spectrum (from discrete landau level to continuous bands) can also be seen in the plot of the energies at $\Gamma$ point in the Inset I. In Fig. 2b, we show an enlarged version of the calculated continuous energy bands. The linear energy dispersion of free massless Dirac Fermions is also imposed for comparison.             

To understand the above-mentioned transition of energy spectrum, we calculated the wavefunctions of different energy levels at $\Gamma$ point. The wavefunction for the Landau level n=4 is plotted in Inset II, where we can see that in this case, the Dirac Fermions is mainly localized inside the magnetic region. The wavefunction is just the usual harmonic oscillator function (Hermite function). For these Landau levels, we can define the 'cyclotron' size of the Dirac Fermion as the distance between two outermost peaks of the Hermite wavefunction. Obviously, the cyclotron size is a function of magnetic field B and the level index n. The size decreases with B and increases with n. The transition of the energy spectrum can then be understood as a consequence of the competition between the cyclotron size of Dirac Fermions and the width of the magnetic region. In the case shown in Fig. 2, for the Landau level n=4, the cyclotron size has been almost the same as the width of the magnetic region (Inset II). When n is bigger than 4, Dirac Fermions escape from the magnetic trap, resulting in the transition of the energy spectrum. Wavefunctions for n=5 and 6 are plotted in Fig. 2c and 2d, respectively. In these cases, the Dirac Fermion is mainly found in non-magnetic regions. For n=5, the wavefunction outside the magnetic trap is similar to the ground state of an electron confined in a box. For n=6, the wavefunction has two peaks, and is analogous to the first excited state of electrons confined in box. The analogy between these Landau-level derived states of massless Dirac Fermions and particle-in-box states of electrons are also correct for other cases with higher n.   

One nice thing of the predicted transition of energy spectrum is that the transition energy where the particle-in-box states occur can be controlled by the magnetic field together with the size of the magnetic region. With the magnetic field of 0.04 $T$, our calculations show that for the case of lattice constant 1500 $nm$ and the magnetic region 375 $nm$, the transition occurs at the $n=1$ Landau level, which means in the energy spectrum, there is only one flat band corresponding to the lowest Landau level ($n=0$). If the size of the magnetic region decreases to 250 $nm$, the lowest Landau level also becomes a particle-in-box state with a continuous energy band. With a weaker magnetic field, the lattice constant and the size of the magnetic region can be bigger for the transitions to occur. These findings offer a new and also practical methodology for controlling/tuning massless Dirac Fermions in graphene that may have great implications for future design of graphene-based devices.

The co-existence of Landau levels and particle-in-box states in the energy spectrum is also observed in ferromagnetic cases or 2d superlattices. We present in Fig. 3 particle-in-box states for a special type of 2d AF superlattices as depicted in Fig. 1b. The wavefunction in Fig. 3a is analogous to the ground state of the particle-in-box states. Fig. 3b and c are similar to the first and second excited states. A very high excited state is plotted in Fig. 3d.

\begin{figure}
	\centering
	\includegraphics[width=12cm]{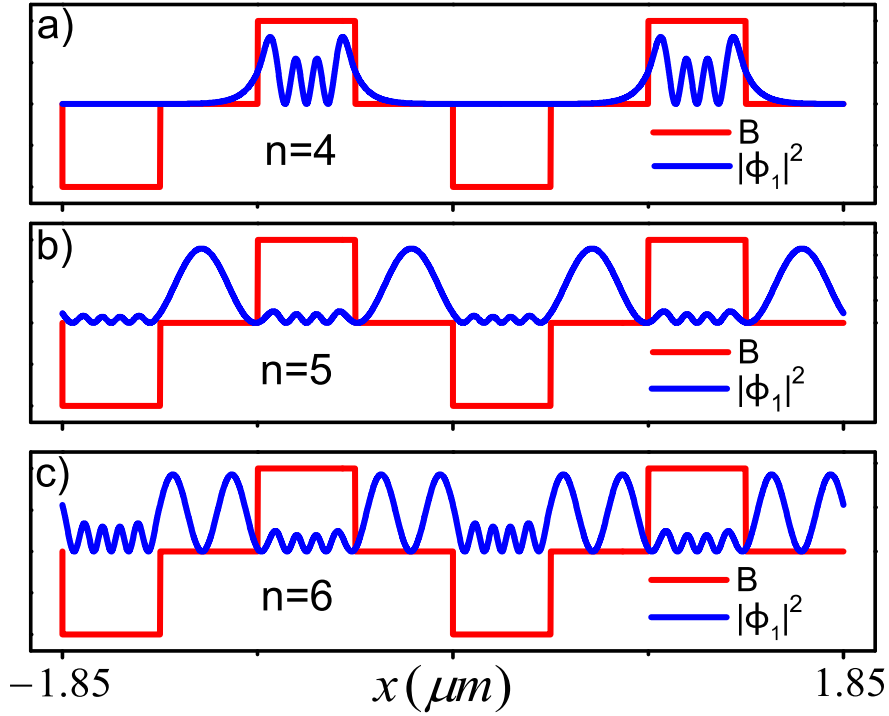}
	\caption{\label{fig:type1} (Color online) First component of pseudo spinor, $|\phi_{1}|^{2}$ for 2d superlattice. The amplitude increases from purple to wine. The black circle indicates the boundary of the magnetic well. The corresponding energies for the four eigen states are $E_{a}=10.1meV$, $E_{b}=16.1meV$, $E_{c}=21.4meV$, $E_{d}=41.4meV$.}
\end{figure} 
 
In summary, by numerically solving the spinor equations, we show that graphene under periodic magnetic field exhibits a unique energy spectrum with the discrete Landau levels at low energies and continuous energy bands at higher energies. The continuous energy bands originate from the wavefunctions analogous to particle-in-box states of electrons. The transition energy where the particle-in-box states occur can be controlled by the magnetic field together with the geometry of the superlattice structures, which offers new avenues for the design of graphene-based systems with unique properties. The magnetic superlattice structures discussed in this paper may be readily fabricated by putting graphene onto a magnetic substrate with pre-designed 1d or 2d periodic surface patterns. The strength of the magnetic field (around 0.1 T) as well as the detailed superlattice structures (for example, 1850 $nm$ of lattice constant) are all within the capabilities of current experiment techniques. We expect our findings to stimulate new experiments along this direction.

The research is supported by NUS research academic fund (Grant no. R144-000-298-112). Computations were performed at the Center for Computational Science and Engineering (CCSE) in NUS.


\end{document}